\documentclass[prm,twocolumn,superscriptaddress, amsmath,amssymb]{revtex4}
\usepackage{graphicx,color,dcolumn,bm,nicefrac}

\newcommand{\vect}[1]{\mathbf{#1}}

\newcommand{\rfm}{$\rm RbFe(MoO_4)_2$}

\begin{document}
\title{High-field magnetic structure of the triangular antiferromagnet RbFe(MoO$_4$)$_2$}

\author{Yu.~A.~Sakhratov}
\affiliation{National High Magnetic Field Laboratory, Tallahassee, Florida 32310, USA}
\affiliation{Kazan State Power Engineering University, 420066 Kazan, Russia}
\author{O.~Prokhnenko}
\affiliation{Helmholtz-Zentrum Berlin f\"ur Materialien und Energie GmbH, Hahn-Meitner-Platz 1 D-14109 Berlin, Germany}
\author{A.~Ya.~Shapiro}
\affiliation{A.V. Shubnikov Institute of Crystallography RAS, 119333 Moscow, Russia}
\author{H.~D.~Zhou}
\affiliation{Department of Physics and Astronomy, University of Tennessee, Knoxville, Tennessee 37996, USA}
\author{L.~E.~Svistov}
\email{svistov@kapitza.ras.ru}
\affiliation{P.L. Kapitza Institute for Physical Problems, RAS, Moscow 119334, Russia}
\author{A.~P.~Reyes}
\affiliation{National High Magnetic Field Laboratory, Tallahassee, Florida 32310, USA}
\author{O.~A.~Petrenko}
\affiliation{Department of Physics, University of Warwick, Coventry CV4 7AL, United Kingdom}

\date{\today}
\begin{abstract}
The magnetic  $H-T$ phase diagram of a quasi-two-dimensional antiferromagnet \rfm\ ($S=5/2$) with an equilateral triangular lattice structure is studied with $^{87}$Rb NMR and neutron diffraction techniques.
This combination of experimental techniques allows us to determine the ordered components of the magnetic moments on the Fe$^{3+}$ ions within various high-field phases -- the Y, UUD, V, and fan structures, stabilized in the compound by the in-plane magnetic field.
It is also established that the transition from the V to the fan phase is of first-order, whereas the transition from the fan phase to the polarized paramagnetic phase is continuous.
An analysis of the NMR spectra shows that the high-field fan phase of \rfm\ can be successfully described by a periodic commensurate oscillation of the magnetic moments around the field direction in each Fe$^{3+}$ layer combined with an incommensurate modulation of the magnetic structure perpendicular to the layers.
\end{abstract}
\maketitle
\section{Introduction} \label{Intro}

\rfm\ is an example of a quasi-two-dimensional (2D) antiferromagnet ($S=5/2$) with a triangular lattice structure.
In this compound the antiferromagnetic inter-planar exchange interaction $J'$ is much weaker than the dominant in-plane interaction, $J$.
For a magnetic field applied within the triangular plane, a sufficiently strong single-ion easy-plane anisotropy leads to a strong resemblance between the main features of the $H-T$ phase diagram for \rfm~\cite{Inami_1996, Svistov_2003, Kenzelmann_2004, Svistov_2006, Kenzelmann_2007, Smirnov_2007, White_2013, Mitamura_2014, Zelenskiy_2021} and the model phase diagram of the $XY$ 2D triangular lattice antiferromagnet (TLAF) studied in detail in Refs.~\cite{Korshunov_1986, Lee_1986, Chubukov_1991,Gekht_1997}.

The magnetic energy of the TLAF can be described by the following expression:
\begin{equation}
E = J\sum_{<i,j>} \mathbf{S}_i \mathbf{S}_j - \sum_i \mathbf{H} \mathbf{M}_i .
\label{eqn_energy}
\end{equation}
Here the sum in the first term is through all the exchange bonds of the nearest magnetic ions of a triangular lattice, $\mathbf{H}$ is external magnetic field, $\mathbf{M}_i=-g\mu_B \mathbf{S}_i$ and $g$ is the g-factor.

 \begin{figure}[tb]
\includegraphics[width=0.85\columnwidth]{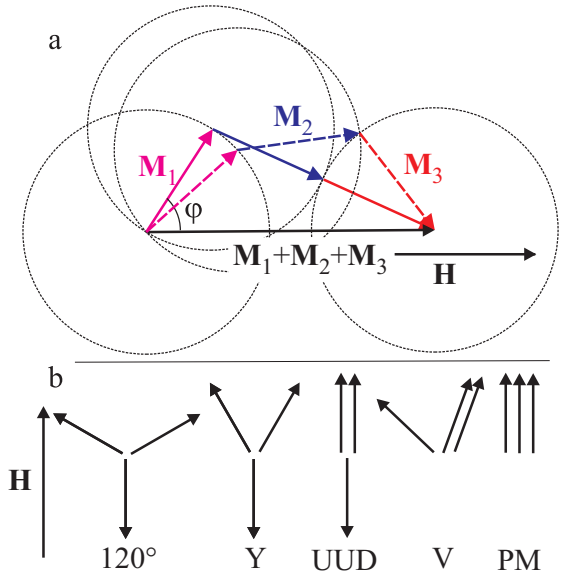}
\caption{(a) Illustration of the spin arrangements for two degenerate magnetic structures selected (in the framework of a classical model) from the manifold of magnetic phases.
Here the total magnetization is arbitrarily set to $0.8M_{\rm sat}$.
The structure represented by the solid arrows corresponds to the V~phase.
(b) Magnetic structures expected for the XY TLAF model which includes spin fluctuations.}
\label{fig1_structures}
\end{figure}

The minimization of the magnetic energy given in Eq.~\ref{eqn_energy} in the classical limit of large spins demonstrates a strong degeneracy -- the manifold of the magnetic phases with equal net magnetic moment have the same energy~\cite{Korshunov_1986, Chubukov_1991}.
This follows from the fact that the exchange and Zeeman terms of Eq.~\ref{eqn_energy} depend only on the total magnetic moment.
Such a degeneracy can be seen by rewriting the exchange term of the equation as $\frac{NJ}{6}(\vect{S}_1+\vect{S}_2+\vect{S}_3)^2+C$, where $N$ is the total number of spins and $C$ is a field-independent constant.
Figure~\ref{fig1_structures}(a) demonstrates the abundance of three-sublattice planar structures with the fixed magnetic moment $\vect{M}=(\vect{M}_1+\vect{M}_2+\vect{M}_3)N/3$, which, in the classical limit, all have equal energy.
 An energy preference for the different structures appears when a contribution from thermal and quantum fluctuations is considered.
 The sequence of magnetic field induced structures expected in the framework of the $XY$-model which includes the fluctuations is shown in Fig.~\ref{fig1_structures}(b)~\cite{Korshunov_1986, Lee_1986, Chubukov_1991}.

 On application of a magnetic field, the zero-field 120-degrees structure is replaced by the so-called Y~phase, for which the angles between the magnetic moments are altered to induce a net magnetization.
 In a field near one third of the saturation field, $H_{\rm sat}/3$, a collinear magnetic structure labeled as the up-up-down (UUD) phase is realized.
 In this field range, the magnetization shows a plateau at $M(H)=M_{\rm sat}/3$.
 On further increase of the applied field up to $H_{\rm sat}$, a canted phase known as the V~phase with two out of three co-aligned magnetic sublattices is expected.
 The magnetic phase with all the moments directed along the applied field at $M(H)>M_{\rm sat}$ is referred to as a \textbf{}polarized paramagnetic (PM) phase.

The energy gain for the phases stabilized by fluctuations is relatively small.
Therefore in the majority of the experimental realizations of the three-dimensional (3D) TLAF the magnetic structure is defined by other interactions, such as additional in-plane or inter-planar interactions.
\rfm\ is an excellent example of a quasi-2D TLAF with the magnetic order controlled by the fluctuations.
The phase boundaries between the PM and the ordered phases, as well as the exact position of the  UUD~phase on the experimental $H-T$ phase diagram~\cite{Svistov_2003, Kenzelmann_2004, Svistov_2006, Kenzelmann_2007, Smirnov_2007, White_2013, Zelenskiy_2021}, are in a good quantitative agreement with the predictions of the 2D XY model studied in Ref.~\cite{Lee_1986}.
The magnetic phases found in \rfm\  in the intermediate field ranges are in agreement with the 2D XY model, whereas the low and high field magnetic phases differ from the model's predictions.
This observation demonstrates the importance of further interactions in defining the magnetic structure at low and high fields.

Here we report on NMR and neutron diffraction studies of \rfm\ in the high field range, $H \geq H_{\rm sat}/3$, for the field applied within the triangular plane.
The transition to the polarized PM phase at low temperature takes place at $\mu_0 H_{\rm sat} \approx 18$~T~\cite{Inami_1996, Smirnov_2007}.
As a result, we are able to elucidate the magnetic $H-T$ phase diagram.
Prior to describing the experimental procedures and presenting our results, in the next section we review the magnetic properties of \rfm, focusing on the sequence of field-induced magnetic structures of this compound and their relation with the geometry of the lattice.

\section{Magnetic interactions and structure of \rfm}

\begin{figure}[tb]
\vspace{0 mm}
\includegraphics[width=0.85\columnwidth]{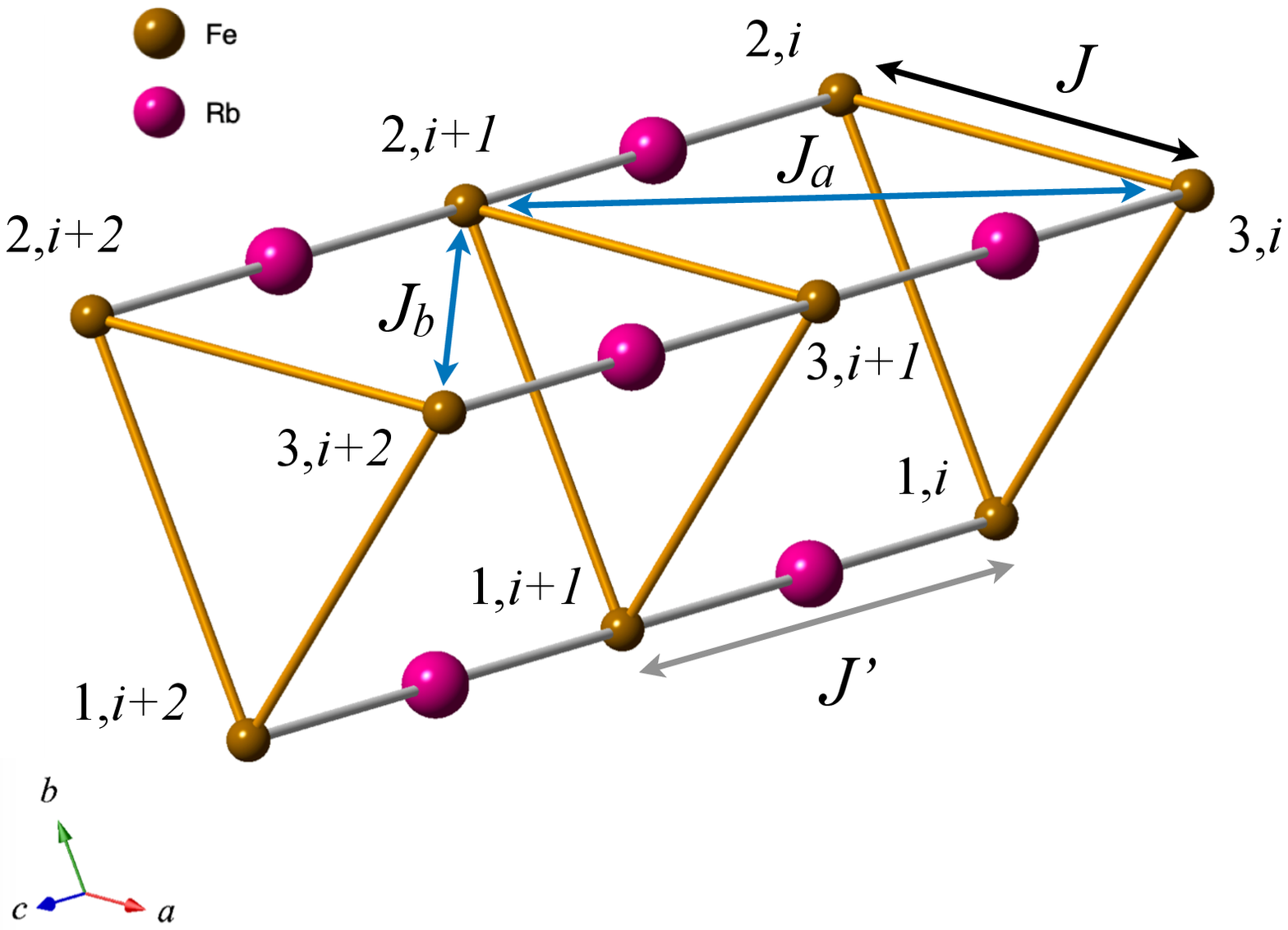}
\vspace{0 mm}
\caption{Crystal structure of \rfm.
The smaller spheres mark the positions of the Fe$^{3+}$ magnetic ions, the larger spheres are at the positions of the Rb$^+$ ions, while the (MoO$_4$)$^{2-}$ complexes are not shown.
The crystal symmetry dictates the following main exchange bonds within one cell: in-plane bonds $J$, inter-plane bonds $J'$, $J_a$, $J_b$.}
\label{fig2_crystal}
\end{figure}

The crystal structure of \rfm\ consists of alternating layers of Fe$^{3+}$, (MoO$_4$)$^{2-}$, and Rb$^+$ ions normal to the threefold axis $C^3$.
Within the layers, the ions form regular triangular lattices.
Figure~\ref{fig2_crystal} shows the magnetic Fe$^{3+}$ ions (5$d^3$, $S=5/2$) and the non-magnetic Rb$^+$ ions, the (MoO$_4$)$^{2-}$ complexes are omitted for clarity~\cite{Klevtsova_1970}.
The lattice parameters of \rfm\ are $a=5.69$~\AA\ and $c=7.48$~\AA.
At room temperature, the crystal structure  belongs to the space group $P\bar{3}m1$ and below 190~K the symmetry is lowered to the space group $P\bar{3}c1$~\cite{Klimin_2003}.
The low-temperature crystal structure allows for the presence of the following exchange bonds, the in-plane interactions, $J=J_{1,i;2,i}=J_{2,i;3,i}=J_{3,i;1,i}$, the inter-plane interactions  $J'=J_{1,i;1,i+1}=J_{2,i;2,i+1}=J_{3,i;3,i+1}$, $J_a=J_{1,i;2,i+1}=J_{2,i;3,i+1}=J_{3,i;1,i+1}$ and $J_b=J_{2,i;1,i+1}=J_{3,i;2,i+1}=J_{1,i;3,i+1}$.
The in-plane exchange integral $J=0.086(2)$~meV~\cite{White_2013} is antiferromagnetic and is much stronger than the inter-planar integrals $J'$, $J_a$, $J_b$.
The inter-plane exchange integral is approximately one hundred times smaller than the in-plane exchange integral, $J'/J=0.008(1)$ according to neutron diffraction experiments~\cite{White_2013} and $J'/J=0.01$ according to ESR measurements~\cite{Svistov_2003,Svistov_2006_Err}.
The ratio of the diagonal inter-plane exchange integrals was also evaluated from first principle calculations~\cite{KunCao_2014} as $|J_{a,b}| / J \approx 0.002$.

\begin{figure}[tb]
\includegraphics[width=0.95\columnwidth]{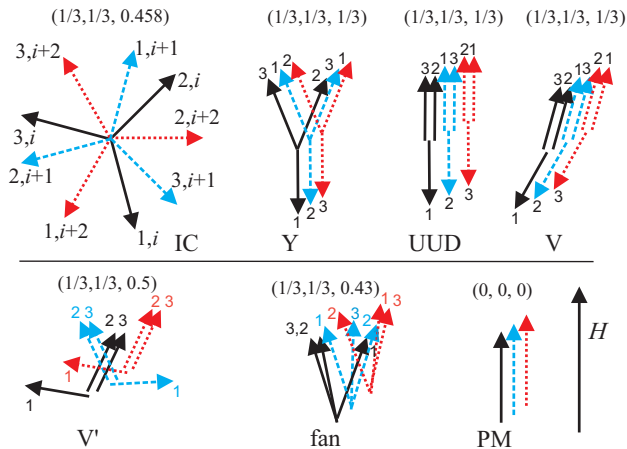}
\caption{Magnetic structures and their propagation vectors \cite{Kenzelmann_2007} expected in \rfm\ for the field $H$ applied within the basal plane.
The arrows show the spin directions of the three neighboring Fe$^{3+}$ ions within a triangular plane.
The spins from three adjacent planes are shown with different colors and line styles.
The sequence of structures shown corresponds to an increasing magnetic field.}
\label{fig3_MagStructures}
\end{figure}

\rfm\ orders magnetically below $T_{\rm N}\approx 3.8$~K~\cite{Inami_1996}.
The magnetic structures found for the in-plane field orientation using a variety of experimental methods~\cite{Inami_1996,Svistov_2003,Kenzelmann_2004,Svistov_2006,Kenzelmann_2007,Smirnov_2007,White_2013,Mitamura_2014,Zelenskiy_2021} are schematically depicted in Fig.~\ref{fig3_MagStructures}, the sequence shown corresponds to an increasing magnetic field.
The arrows show the spin directions for the three neighboring Fe$^{3+}$ ions forming a triangle in the basal plane, the spins from the three different planes are represented with different colors and line styles.
In zero field, a 120$^{\circ}$ magnetic structure is established within the triangular plane.
Such a structure corresponds to the minimum of the exchange energy within each triangular plane for an antiferromagnetic in-plane exchange interaction.
The much weaker inter-plane interactions stabilize an incommensurate magnetic structure with the propagation vector (1/3,~1/3,~0.458)~\cite{Kenzelmann_2007}.
The wave number of 0.458 reflecting the modulation along the $c$~axis, corresponds to the pitch angle of the neighbor spins from nearest triangular planes of $\approx 165^\circ$, which indicates that the main inter-plane exchange interaction is also antiferromagnetic.

The applied in-plane field disturbs a perfect 120-degree structure in every plane.
The spin fluctuations provide an energy gain for the Y phase [see Fig.~\ref{fig1_structures}(b)], the gain increases with the increasing field from its zero value at $H=0$.
Three consecutive commensurate magnetic phases with the propagation vector (1/3, 1/3, 1/3) are observed within the intermediate field range, $H_{\rm sat}/5\lesssim H \lesssim 2H_{\rm sat}/3$.
The phases, Y, UUD and V shown in Fig.~\ref{fig1_structures}(b) have a favorable spin arrangement within each plane when the spin fluctuations are taken into account.
For these three phases, an alternation of the neighboring planar structures along the $c$~axis is defined by the inter-planar interactions.
Figure~\ref{fig3_MagStructures} depicts the magnetic structure and explicitly labels the 9 spins in the extended magnetic unit cell for these 3D commensurate phases.

These phases are energetically favorable for a model with an antiferromagnetic in-plane exchange interaction $J$ and an antiferromagnetic inter-plane exchange interaction $J'$ between the nearest magnetic ions of the neighboring planes, as considered in Ref.~\cite{Gekht_1997}.
According to this model, a transition to an additional 3D phase V$'$ is expected at a field of $\mu_0 H_c = \mu_0 H_{\rm sat}/\sqrt{3}\approx 10.4$~T (see Appendix~\ref{AppA} for details).
The propagation vector of the V$'$ phase is (1/3, 1/3, 1/2).
Experimental studies of the low temperature magnetic properties of \rfm\  have indeed found a magnetic transition in a field close to $H_c$~\cite{Svistov_2006, Kenzelmann_2007, Smirnov_2007, White_2013, Zelenskiy_2021}, however,  contrary to the expectations from the model, the magnetic structure observed in the high-field range, $H_c<H<H_{\rm sat}$ is incommensurate, with a propagation vector (1/3, /1/3, 0.43)~\cite{Kenzelmann_2007}.

A classical consideration of a screw-type magnetic system in Ref.~\cite{Nagamiya_1967} leads to a conclusion that for a field applied within the easy-plane and sufficiently close to the saturation field, an incommensurate planar fan spin structure should be expected.
For the high-field phase ($H \gtrsim 0.72 H_{\rm sat}$) in \rfm\ one could expect the 3D fan structure with the following in-plane spin configuration~\cite{Cemal_2017, Utesov_2020}:
\begin{eqnarray}
\label{eqn_fan}
M_{1,i,x} & = & \mu L \cos(k_{ic}~c~i), 		\nonumber \\
M_{1,i,z} & = & \sqrt{\mu^2-M_{1,i,x}^2}, 		\nonumber \\
M_{2,i,x} & = & \mu L \cos(k_{ic}~c~i+2\pi/3), 	\nonumber \\
M_{2,i,z} & = & \sqrt{\mu^2-M_{2,i,x}^2},				\\
M_{3,i,x} & = & \mu L \cos(k_{ic}~c~i+4\pi/3),	\nonumber \\
M_{3,i,z} &= & \sqrt{\mu^2-M_{3,i,x}^2},		\nonumber
\end{eqnarray}
here the field is applied along the $z$~axis,
$k_{ic}=0.43 \times 2\pi/c$, $i$ is the index of triangular plane, 1, 2, and 3 are the indices of the in-plane magnetic sublattices, $\mu$ is the magnetic moment on the Fe$^{3+}$ ion, and $L$ is a field dependent order parameter, $L<1$, that becomes zero at $H_{\rm sat}$.
The analysis of the fan structure shows that the total magnetic moment of every plane is directed along the applied field and its value oscillates from one plane to another around the mean value of $M_0=\sum\limits_{i=1,N}(M_{1,i}+M_{2,i}+M_{3,i})/3N$.
An oscillation of the magnetic moments of stacked triangular planes leads to the variation of the magnetic energy (exchange and Zeeman) as discussed in the Sec.~\ref{Intro}.
In the high-field range, the energy variation is perfectly harmonic which leads to the situation where the fan structure turns out to be degenerate with the energetically preferable structures composed of the planes with equal magnetic moments $M_0$ in each plane.
Therefore the energy difference between the structure composed from the manifold of phases with the same $M_0$ magnetic moment in each plane [Fig.~\ref{fig1_structures}(a)] and the fan~structure with the mean magnetic moment $M_0$ practically disappears in a high field range, $0.75 M_{\rm sat}<M_0<M_{\rm sat}$.
More detailed discussion of the fan structure is given in Appendix~\ref{AppB}.
This calculation suggests that the fan magnetic structure might be stabilized by an interaction much weaker than the main in-plane exchange $J$.
Possible candidates for such an interaction are the weak diagonal inter-plane exchange interactions, $J_a$ and $J_b$ [see Fig.~\ref{fig1_structures}(b)].

Concluding the description of the magnetic phases of \rfm, we note that the magnetic moments of the magnetic ions consist partly of magnetically ordered components discussed above, and partly of the moments of the field-polarized fluctuating spins.
The first component acts as an order parameter and decreases with increasing temperature on approach to the PM phase in contrast to the second component.
The magnetic phases of \rfm\ were studied at several positions on the $H-T$ phase diagram (Fig.~\ref{fig4_PhD}) by neutron diffraction experiments~\cite{Kenzelmann_2007, White_2013, Mitamura_2014}.
The study of the 120$^{\circ}$~phase at $H=0$, $T=2$~K shows that the value of the ordered component of magnetic moment is 3.9(5)$\mu_{\rm B}$, or 0.78$g\mu_{\rm B}S$.
For the UUD~phase studied at $\mu_0 H=6$~T, $T=2$~K, the ordered component of the magnetic moment is 3.3(3)$\mu_{\rm B}$~\cite{Kenzelmann_2007} or 0.66$g\mu_{\rm B}S$.
Combining these results with the magnetic moment measured at the same conditions, $M=1.6(5)\mu_{\rm B}/{\rm Fe}^{3+}$~\cite{Svistov_2003}, the uniform magnetization can be evaluated as 0.5$\mu_{\rm B}/{\rm Fe}^{3+}$ or 0.1$g\mu_{\rm B}S$.
The V~phase (Fig.~3) was also studied at $\mu_0 H=10$~T, $T=0.5$~K.
At this point the ordered component of magnetic moment is 3.7(5)$\mu_{\rm B}$~\cite{Kenzelmann_2007}, which is close to the value of the ordered component in the UUD~phase.
The uniform magnetization can be evaluated as 0.3$\mu_{\rm B}$.
The neutron diffraction study of the high field phase at $\mu_0 H=14.9$~T and $T=100$~mK ~\cite{Kenzelmann_2007, White_2013} indicates a magnetic structure which is more complicated than the fan structure described by Eq.~\ref{eqn_fan}.
According to this study, the projections of the magnetic moments to the axis perpendicular to the applied field varies in accordance with Eq.~\ref{eqn_fan}, with $\mu L= 1.88\mu_{\rm B}$, while the projections of the magnetic moments in the direction of applied field, 3.8~$\mu_{B}$, is the same for all Fe$^{3+}$ ions.
Such a structure implies a spatial modulation of the magnetic moment values with an amplitude of $\approx 0.2 \mu_{\rm B}$.

In this paper we report on NMR and neutron diffraction measurements performed on single crystal samples which allow us to clarify the nature of the Y, UUD, V and fan magnetic phases of \rfm.
The studies permit us to obtain the ordered components of magnetic ions within these phases.
It is also established that the transition from the V to the fan phase is of the first-order type, whereas the transition from the fan phase to the polarized PM phase is continuous.
This observation excludes the additional high field  phase transitions suggested in Refs.~\cite{Korshunov_1986, Smirnov_2007}.
The analysis of the NMR spectra shows, that the high field fan structure of \rfm\ can be fully described by Eq.~\ref{eqn_fan}.

\section{Sample preparation and experimental details}

We used two different batches of \rfm\ single crystals for the measurements.
The samples from the first batch (labeled in the following as batch~I) were the same as used in Refs.~\cite{Svistov_2006, Smirnov_2007}.
The second batch was prepared at the University of Tennessee using the technique described in Ref.~\cite{Svistov_2003}.
A powder x-ray diffraction study of the structures did not show a significant difference between the batches, however, the magnetic ordering temperature $T_{\rm N}$ and the values of $H_{\rm sat}$ for the second batch were about 7\% lower than for the samples from the first batch.
The positions of the phase transition and the shapes of NMR spectra for both set of samples were nearly the same when scaled by $T/T_{\rm N}$ and $H/H_{\rm sat}$.
A typical size of the crystal was $2 \times 2 \times 0.5$~mm$^3$ with the smallest dimension corresponding to the $C^3$ direction of the crystal.

NMR measurements were taken in a superconducting Cryomagnetics 17.5~T magnet at the National High Magnetic Field Laboratory, Florida, USA.
$^{87}$Rb nuclei (nuclear spin $I=3/2$, gyromagnetic ratio $\gamma/2\pi=13.9318$~MHz/T) were probed using a pulsed NMR technique.
The spectra were obtained by summing fast Fourier transforms (FFT) spin-echo signals as the field was swept through the resonance line.
Utilizing FFT techniques, NMR spin echoes were obtained using $\tau_p - \tau_D - 2\tau_p$ pulse sequences, where the pulse lengths $\tau_p$ were 1~$\mu$s and the times between pulses $\tau_D$ were 15~$\mu$s.
$T_1$ was extracted using a multiexponential expression which is utilized in spin-lattice relaxation when NMR lines are split by quadrupole interaction~\cite{Suter_1998}.
The measurements were carried out in the temperature range $1.43 \leq T \leq 25$~K, temperature stability was better than 0.05~K.

\begin{figure}[tb]
\includegraphics[width=0.95\columnwidth]{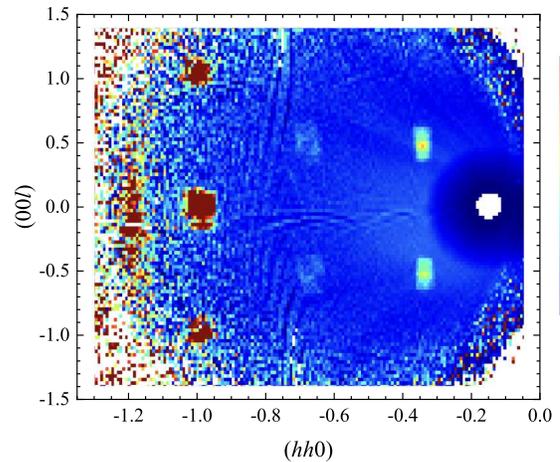}
\caption{Neutron diffraction intensity map of the $(hhl)$ plane measured in \rfm\ single crystal sample at $T=1.4$~K in a field of 3~T applied perpendicular to the $c$~axis.
Both nuclear reflections $(\bar{1}\bar{1}\bar{1})$,  $(\bar{1}\bar{1}0)$, $(\bar{1}\bar{1}1)$ and magnetic reflections $(\bar{\frac{1}{3}}\bar{\frac{1}{3}} q_z)$, $(\bar{\frac{2}{3}}\bar{\frac{2}{3}} q_z)$ with $q_z \approx \pm \frac{1}{2}$ are captured within the covered reciprocal space.}
\label{fig_N1}
\end{figure}

The neutron scattering experiment has been performed at the HFM/EXED high-field facility at Helmholtz-Zentrum Berlin, Germany.
At the end of 2019 the facility stopped its operation because of the planned shutdown of the BER II research reactor.
Until that time, it was the only place in the world where neutron scattering experiments in continuous magnetic fields up to 26~T could be performed.
The facility combined the horizontal field hybrid magnet (HFM) and the time-of-flight EXtreme Environment Diffractometer (EXED) \cite{Prokhnenko_2017,Smeibidl_2016,Prokhnenko_2016}.
The HFM had 30~degrees conical openings on both ends of a 50~mm diameter room-temperature bore and could be rotated with respect to the incident neutron beam by an angle of up to 12 degrees.

In the present experiment, the sample was aligned such that the scattering plane was spanned by the vectors [-110] and [110].
The magnetic field, the direction of which was kept fixed during the experiment, was oriented along the former; the magnet was rotated to -11.8~degrees.
An experimentally determined sample misalignment around the field axis was 3~degrees.
Typical exposure times were 1 to 2 hours per temperature or field.
The sample was mounted in a $^4$He flow cryostat with base temperature of about 1.4~K.
The cryostat was equipped with a sample rotation stage around the vertical axis.
A combination of the 30~degrees detector coverage in forward scattering with a wavelength range of $0.7-3.8$~\AA\ allowed a $Q$-range visualized in Fig.~\ref{fig_N1} to be covered.
Each of the magnetic reflections seen at the $(\bar{\frac{1}{3}}\bar{\frac{1}{3}} q_z)$ and $(\bar{\frac{2}{3}}\bar{\frac{2}{3}} q_z)$ positions with $q_z \approx \pm \frac{1}{2}$, consist of two incommensurate peaks, $q_z=0.458$ and $q_z=0.542$, as observed in previous neutron diffraction experiments~\cite{Kenzelmann_2007, White_2013, Mitamura_2014}, however, the $Q$-resolution of the EXED instrument in forward scattering did not allow us to fully resolve the peaks.

\begin{figure}[tb]
\includegraphics[width=0.95\columnwidth]{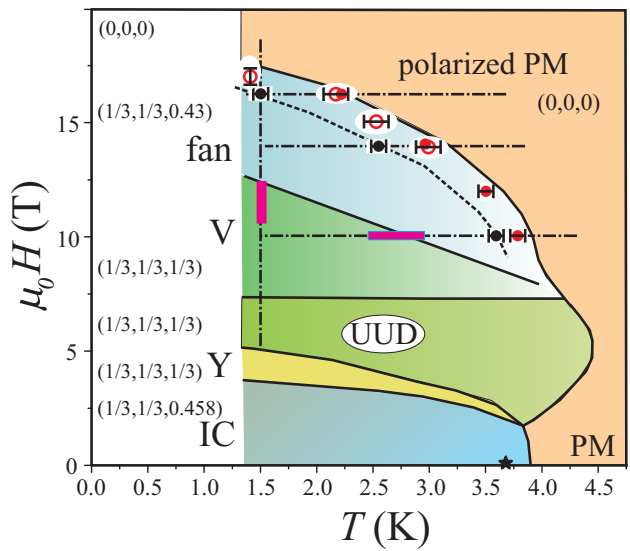}
\caption{Magnetic phase diagram of \rfm, the magnetic phases are labeled in accordance with the notations used in Fig.~\ref{fig3_MagStructures}.
The open red circles mark the transitions to the PM phase obtained from the neutron diffraction experiments on a sample from batch I.
The solid circles show the transitions to the PM phase obtained from the NMR experiments on samples from batches I and II, the symbols are colored red and black, respectively.
A black star shows the $T_{\rm N}$ measured in samples of batch II from the temperature dependence of the magnetization at $\mu_{0}H=0.1$~T.
The solid black lines show the phase boundaries obtained with different experimental technique from Ref.~\cite{Smirnov_2007} on samples from batch I.
Dash-dotted lines show the temperature and field scans made in the NMR study.
The wave vectors of the magnetic phases are given from Ref.~\cite{{Kenzelmann_2007}}.
The vertical and horizontal pink bars mark the transition regions where the magnetic structure is changing from one phase to another. }
\label{fig4_PhD}
\end{figure}

The magnetic $H-T$ phase diagram of \rfm\ is shown in Fig.~\ref{fig4_PhD}.
The dash-dotted lines represent the temperatures and fields at which the NMR and neutron diffraction studies were carried out.
The solid black lines show the phase boundaries obtained with different experimental techniques and collected in Ref.~\cite{Smirnov_2007} on the samples from batch I.
The neutron diffraction experiments at fields of 14, 15, 16, and 17~T described below show that in the high field range, the integrated intensity of the magnetic reflections corresponding to the magnetic structure with the propagation vector $(1/3, 1/3, 0.43)$ tends to zero at the position $T_{\rm N}(H)$ shown in the phase diagram with the open circles.
The solid red circles represent the transition points obtained from a sharp anomaly of the NMR spin-lattice relaxation time as described below.
The phase boundary between the ordered and PM phases obtained from the neutron diffraction and NMR experiments on the samples from the same batch I overlap well.
This allow us to consider in a later discussion of the high-field NMR experiments, only the high-field magnetic structures with the incommensurate wave vector $(1/3, 1/3, 0.43)$.

\section{Experimental results}

\begin{figure}[tb]
\includegraphics[width=0.8\columnwidth]{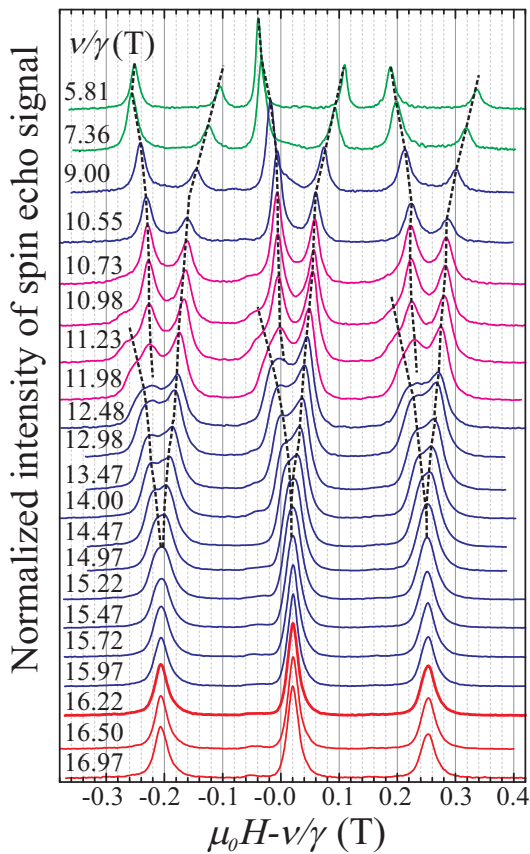}
\caption{$^{87}$Rb NMR spectra measured at $T=1.45$~K in the range from $\mu_0 H_{\rm sat}/3$ to 17~T with the field applied perpendicular to $c$~axis of the crystal on the samples from batch II.
Lines measured at different frequencies are offset for clarity.
Red lines are the spectra measured in a field-polarized PM phase.
Blue lines are the spectra obtained in fields below $H_{\rm sat}$ where the fan phase is expected.
The spectra in low fields shown in green are obtained in the commensurate phases, UUD or V.
Magenta lines show the spectra within the field range of the transition between the commensurate and incommensurate phases.
Dashed lines are guides to the eye.}
\label{fig5_NMR_all}
\end{figure}

\begin{figure}[tb]
\includegraphics[width=0.95\columnwidth,angle=0,clip]{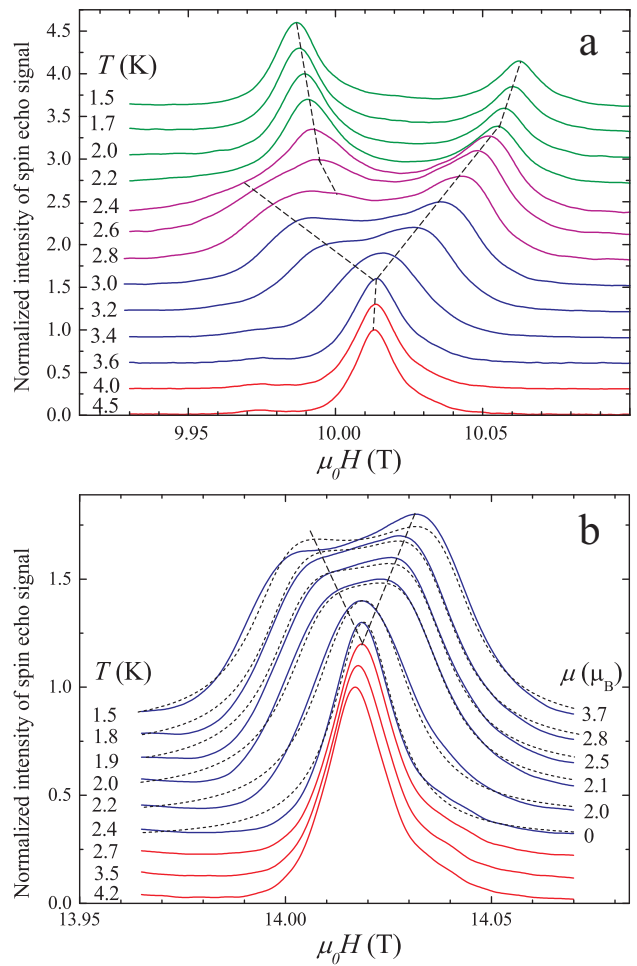}
\caption{Temperature evolution of the $^{87}$Rb NMR spectra measured at (a) $\nu =139.3$~MHz  ($\nu/\gamma \approx 10$~T) and (b) $\nu =195$~MHz ($\nu/\gamma \approx 14$~T).
Dashed lines are guides to the eye.
Colors are used to indicate the spectra obtained in different magnetic phases, PM (red), fan (blue), V (green), and intermediate (magenta).
The dotted lines are the NMR-spectra computed for the fan structure model with the magnetic moments of Fe$^{3+}$ ions shown in (b) near the spectra.
The sample used is from batch~II.}
\label{fig6_NMR_10_14T}
\end{figure}

\begin{figure}[tb]
\includegraphics[width=0.95\columnwidth]{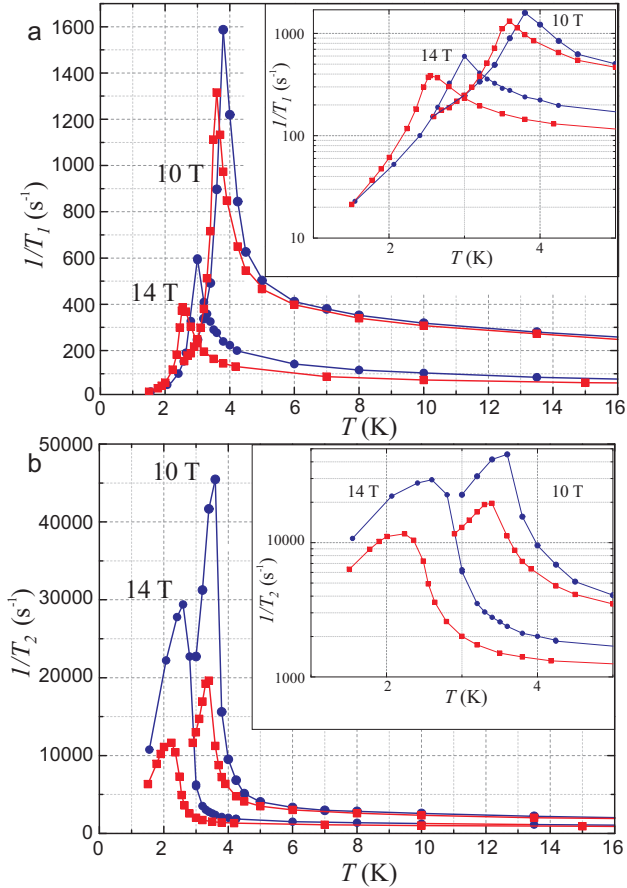}
\caption{Temperature dependences of (a) the spin-lattice, $1/T_1$, and (b) the spin-spin,  $1/T_2$, relaxation frequencies.
The measurements were taken on the central line of the spectrum at $\nu =139.3$~MHz, $\mu_0H=\nu/\gamma \approx 10$~T and $\nu =195$~MHz, $\mu_0H=\nu/\gamma \approx 14$~T.
The points corresponding to the measurements on the sample from batches I and II are shown with the blue circles and red squares, respectively.
The insets show the same dependencies on a logarithmic scale.}
\label{fig7_NMR_T1T2}
\end{figure}

Figure~\ref{fig5_NMR_all} shows the $^{87}$Rb NMR spectra measured at $T = 1.45$~K in the field range $\mu_0 H_{\rm sat}/3 < \mu_0 H <17$~T with the field applied perpendicular to $c$~axis of the crystal.
In the polarized PM phase, all the Rb$^+$ ions are in equivalent positions and the spectrum consists of three lines, the central line corresponds to the transition $m_I = -1/2 \leftrightarrow +1/2$ and the two quadrupole split satellites correspond to the transitions $m_I = \pm 3/2\leftrightarrow \pm 1/2$.
In the magnetically ordered phase, the positions of the Rb$^+$ ions are not equivalent and, as a result, the local magnetic fields from the magnetic neighbors are different and each line of the quadrupole split spectra at $H <H_{\rm sat}$ would exhibit a complex structure.
The effective fields from the ordered components of the magnetic moments are smaller than the field difference for the quadrupolar satellites, and that allows for the observation of the fine structure on every satellite without interference.
The obtained spectra are shown in Fig.~\ref{fig5_NMR_all}  against $\mu_0 H-\nu/\gamma$, where $\nu$ is frequency.
Neglecting the variation of the spin-spin relaxation time, $T_2$, within the NMR spectrum, the measured intensity of the echo signal is proportional to the number of rubidium nuclei in the effective field from the magnetic neighbors equal to $\nu/\gamma-\mu_0 H$.
The field evolution of the NMR spectra within the studied field range can be provisionally divided into three regions, fields above $H_{\rm sat}/3$ but lower than 11~T, fields between 11~T and $H_{\rm sat}$, and fields above $H_{\rm sat}$.
For $T=1.45$~K, the transition from the polarized PM phase to the magnetically ordered phase at $\mu_0 H_{\rm sat} = 16.25$~T was found from the lambda-anomaly of the field dependence of the spin lattice relaxation time $T_{1}$.
The spectra obtained in the PM phase in Figs.~\ref{fig6_NMR_10_14T} and \ref{fig7_NMR_T1T2}, are marked in red.

With decreasing field, the signal initially broadens and then transforms to a spectrum with two maxima.
The intensity of the higher-field signal is slightly higher than the intensity of the low-field signal (see the spectra in blue in Fig.~\ref{fig5_NMR_all}).
The spectra obtained within the field range $\mu_0H_{\rm sat}/3<\mu_0H<11$~T have the same structure as the spectra obtained in Ref.~\cite{Svistov_2005} in a field close to $H_{\rm sat}/3$.
For this field range, the intensity ratio for the two maxima is 2 to 1.
A transformation from the high-field shape to the low-field shape of the spectra takes place in a rather broad field range, 10.7 to 12.5~T, where the features from both shapes coexist.
These spectra are shown in magenta in Fig.~\ref{fig5_NMR_all}.

The temperature evolution of the $^{87}$Rb NMR spectra measured at $\nu =139.3$ and 195~MHz ($\nu/\gamma \approx 10$ and 14~T) are shown in Fig.~\ref{fig6_NMR_10_14T}.
Colors are used to mark the spectra obtained in different magnetic phases, PM (red), fan (blue),  V (green), and mixed (magenta) phases.
We only shown the central line of the spectra corresponding to the transition $m_I = -1/2\leftrightarrow +1/2$.

\begin{figure}[tb]
\includegraphics[width=0.95\columnwidth]{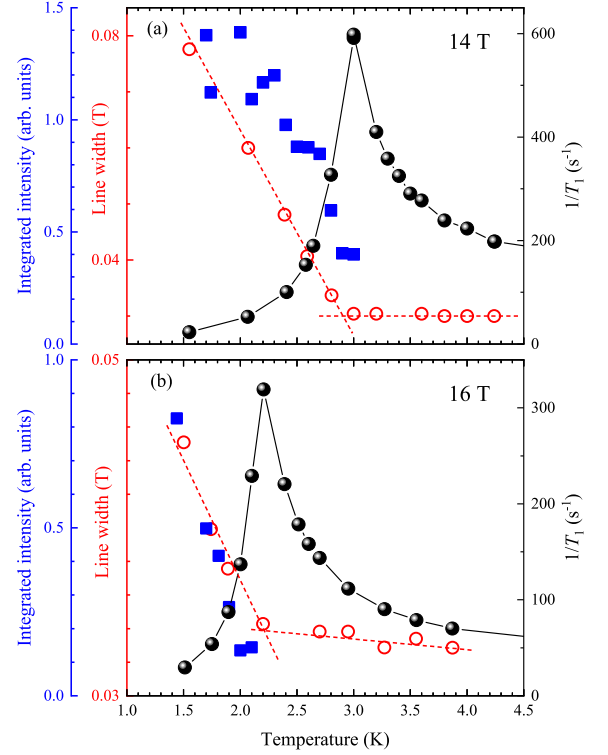}
\caption{Temperature dependence of $1/T_{1}$ and full width at half maximum of the NMR line (circles) at (a) 14 and (b) 16~T.
The square symbols show the temperature dependence of the integrated intensity of (1/3,1/3, $\approx$0.5) neutron diffraction peak measured at the same fields.
The samples used for the measurements were from batch I.}
\label{fig8_Neutr_NMR}
\end{figure}

Figure~\ref{fig7_NMR_T1T2} shows the temperature dependences of the spin-lattice and spin-spin relaxation frequencies, $1/T_1$ and $1/T_2$, measured on the central line of the spectra at 10 and 14~T on the samples from batches I and II.
The temperature dependence of $1/T_1$ demonstrate a sharp lambda-shaped anomaly which coincides with the temperature at which the NMR line broadens.
We associate  this temperature with the transition from a disordered PM phase to a magnetically ordered phase.
The NMR line broadening at the ordering temperature is due to the magnetic fields on the rubidium ions from the ordered moments of the neighboring Fe$^{3+}$ ions.
Figure~\ref{fig8_Neutr_NMR} shows the temperature dependence of $1/T_1$  as well as the full width (half maximum) of the NMR line  obtained in an applied field of 14~T (top panel) and 16~T (bottom panel) for the sample from batch I.

Figure~\ref{fig8_Neutr_NMR} also combines the results of the neutron diffraction experiments on the structure with (1/3, 1/3, $\approx$1/2) propagation vector with the NMR results.
Within the experimental accuracy, the intensity of the magnetic reflections tends to vanish at the temperature of the lambda anomaly in $1/T_1$ which also coincides with the onset of line broadening for both values of the applied field.
This result allows us to limit the discussion of the high field phase to only the structures with the (1/3, 1/3, $\approx$1/2) propagation vector.
We note here that at temperatures above the lambda anomalies $T_1(T)$, $T_2(T)$ and the linewidths do not demonstrate any singularities.
This result shows that there are no other high field phase to be considered.

\begin{figure}[tb]
\includegraphics[width=0.75\columnwidth,angle=270]{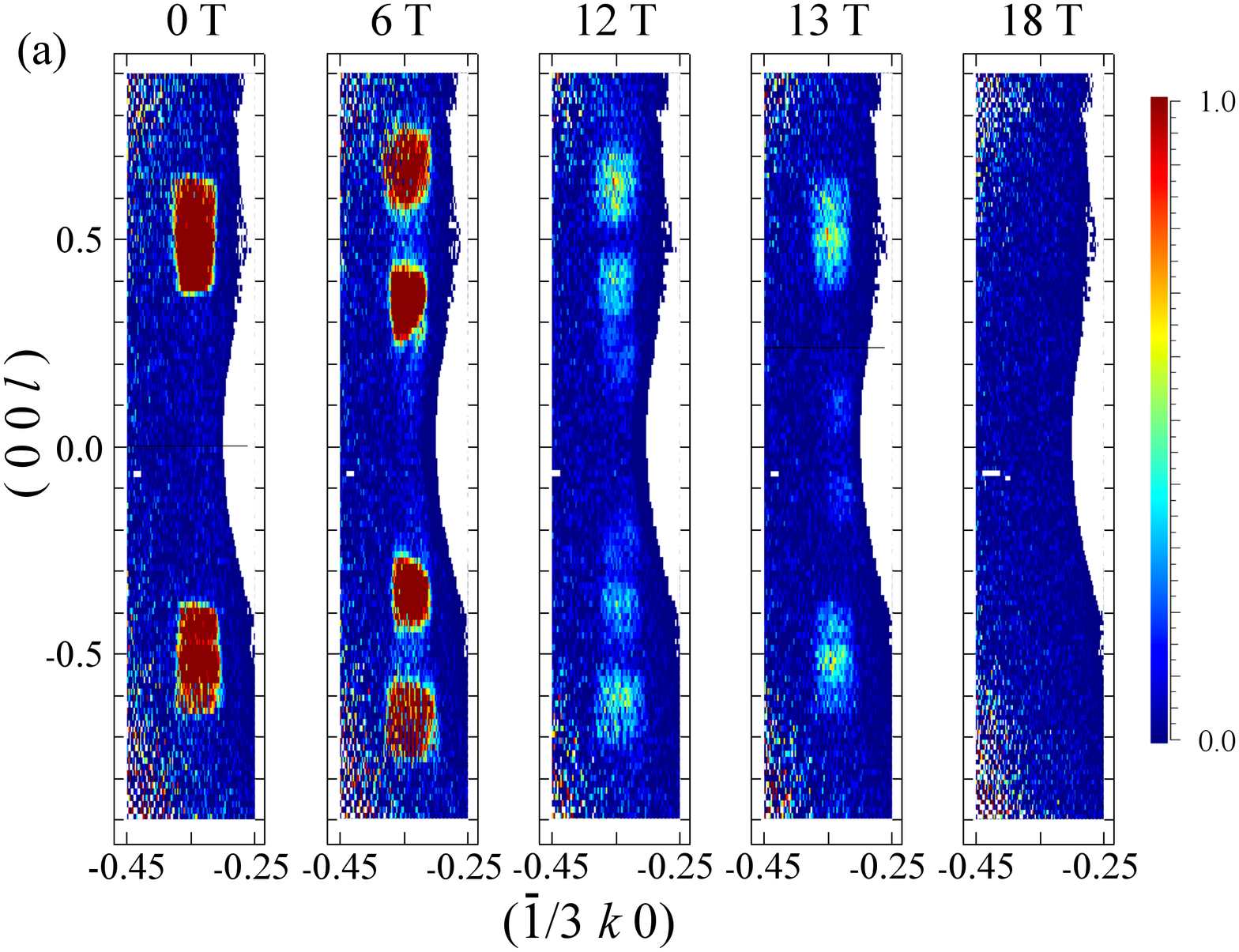}
\includegraphics[width=0.95\columnwidth]{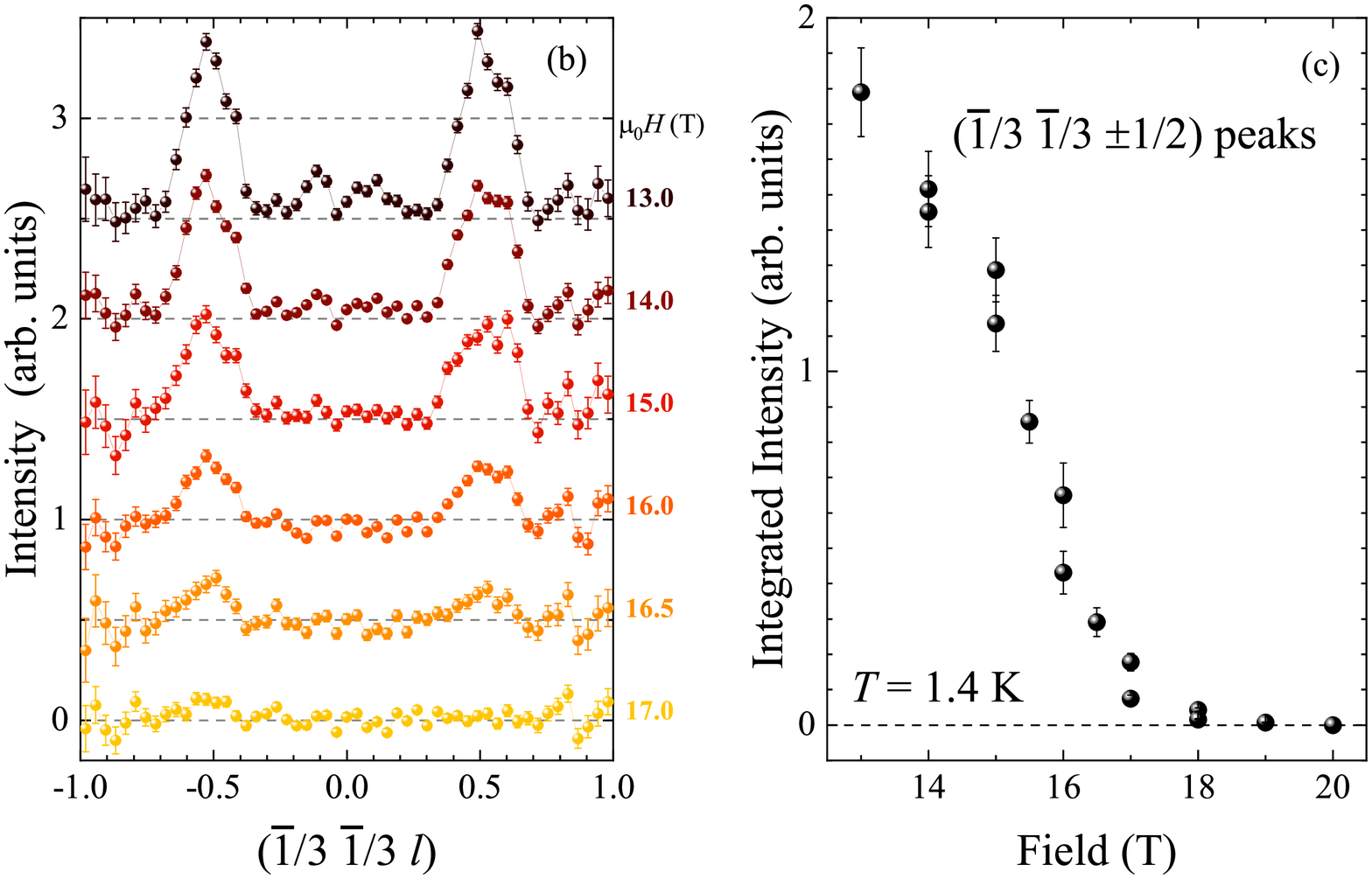}
\caption{(a) Magnetic neutron diffraction intensity maps of the $(0kl)$ plane observed in \rfm\ single crystal sample in different fields applied perpendicular to the $c$~axis.
The magnetic component was isolated by subtracting a high-field background signal corresponding to the polarized PM phase.
(b) Field evolution  between 13 and 17~T of the magnetic intensity for the reciprocal space cuts along the $c^*$ axis taken at the $h=k=\bar{1}/3$ positions.
(c) Field dependence of the total magnetic intensity averaged over the $(\bar{1}/3~\bar{1}/3~1/2)$ and $(\bar{1}/3~\bar{1}/3~\bar{1}/2)$ peaks.
For all three panels the data shown are measured at $T=1.4$~K.}
\label{fig_N2}
\end{figure}

Figure~\ref{fig_N2}(a) follows the field evolution of the magnetic neutron diffraction intensity observed in the $(\bar{1}/3~k~l)$ scattering plane at 1.4~K.
As the maps show nearly perfect symmetry for the positive and negative values of $l$, we therefore describe below only positive $l$.
In zero field, the incommensurate reflections were reported at the $(1/3~1/3~0.458)$ and $(1/3~1/3~0.542)$ positions~\cite{Kenzelmann_2007, White_2013, Mitamura_2014}.
They partially overlap in our measurements and form a single peak seen at $(\bar{1}/3~\bar{1}/3~1/2)$.
For the UUD structure in 6~T, the intense magnetic peaks are clearly seen at the $(\bar{1}/3~\bar{1}/3~1/3)$ and $(\bar{1}/3~\bar{1}/3~2/3)$ positions.
The field of 12~T corresponds to a transition state, where there is a mixture of Bragg peaks from different magnetic phases.
In fact, repeated measurements in this field have shown that the exact ratio of the intensity of the peaks from different structures is very sensitive to sample's history.
In a field of 13~T and above, the history effects seem to be  less important, the diffraction patterns systematically reveal the presence of (structured) peaks near the $(\bar{1}/3~\bar{1}/3~1/2)$ position, composed of the $(1/3~1/3~0.43)$ and  $(1/3~1/3~0.57)$ incommensurate reflections seen in Refs.~\cite{White_2013, Mitamura_2014}.
Apart from the main $(\bar{1}/3~\bar{1}/3~1/2)$ peaks, extra much weaker intensity could also be seen near the $(\bar{1}/3~\bar{1}/3~0.11)$ positions.

Figure~\ref{fig_N2}(b) details the cuts though the magnetic $(\bar{1}/3~\bar{1}/3~l)$ peaks taken along the $l$ direction for the field between 13 and 17~T.
The intensities of the main $(\bar{1}/3~\bar{1}/3~1/2)$ peaks gradually decrease with the applied field, as captured by Fig.~\ref{fig_N2}(c).
The magnetic peaks are completely absent in a field of 18~T and above.
Extra intensity around the $(\bar{1}/3~\bar{1}/3~0.11)$ positions is visible in 13, 14, and perhaps 15 T, while in 16 T and above their intensity does not exceed the noise.
This intensity could be attributed to the presence of a second-order harmonic, reflecting a rather complex modulation of the stacking along the $c$~axis as given by Eq.~\ref{eqn_fan}.
For the high-field phase previously investigated at 14.9~T and 0.1~K by single crystal neutron diffraction, the best fit was obtained with a propagation vector $(1/3~1/3~0.442)$~(see Supplemental Material in Ref.~\cite{White_2013}).
Therefore a second-order harmonic is expected at $(1/3~1/3~0.114)$, exactly where we see an extra intensity in our measurements.
The observation of a second-order harmonic within the fan structure is unsurprising.
According to Eq.~\ref{eqn_fan} the spin components of the magnetic sublattices perpendicular to the applied field change from plane to plane with wave number $k_{ic}$, whereas the components parallel to $H$ oscillate on the even harmonics of $k_{ic}$.
The situation is similar to oscillations of a simple pendulum moving in a vertical gravitational field for which the oscillation projected along the vertical axis take place at the double frequency of the oscillations projected into the horizontal direction.
The appearance of even-order magnetic harmonics in the fan-like structures was reported for several systems~\cite{Givord_1974,Kosugia_2003,Wang_2021}

\section{Discussion}

The magnetic resonance condition of $^{87}$Rb nuclei in our experiments is defined by an interaction with a strong external magnetic field $\nu=\gamma H$, with small corrections due to the quadrupole interaction, dipolar interaction with the magnetic environment and transferred hyperfine interactions with the nearest magnetic neighbors.
The corrections to the resonance field due to a quadrupolar interaction are the same for all $^{87}$Rb nuclei and, as a result, they lead to a field shift of the NMR spectra.
In a high-field range, this shift for the central NMR line $m_I = -1/2\leftrightarrow +1/2$ is negligibly small~\cite{Svistov_2005}.
The largest contribution to the effective field is dipolar, which can be directly computed for the different magnetic structures.

\begin{figure}[tb]
\includegraphics[width=0.9\columnwidth]{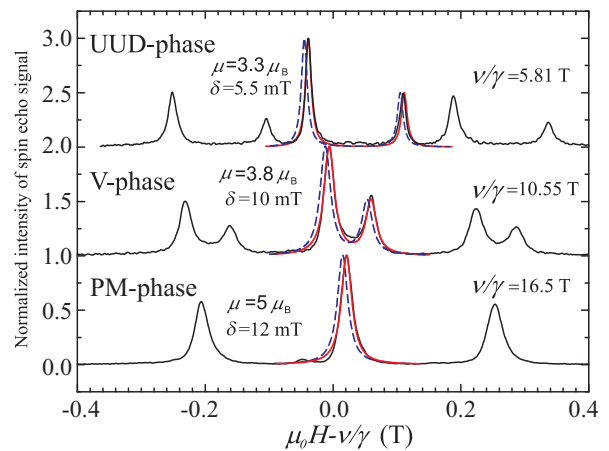}
\caption{Rb NMR spectra measured at $T=1.5$~K in the UUD, V, and PM phases in the sample from batch II  (black lines).
Blue dashed lines show the computed central lines of spectra for the UUD, V and PM structures with the values of the ordered moments on the Fe$^{3+}$ ions, $\mu$, and the full width at half maximum of the individual NMR lines, $\delta$, stated for each spectrum.
These parameters provide a good match to the shape of the experimental spectra.
The red lines show the computed spectra shifted by approximately 0.01 T so that they best coincide with the positions of the experimental lines.}
\label{fig9_spectra_fits}
\end{figure}

The dipolar fields were calculated for a $^{87}$Rb nucleus located in the middle of a cylindrical sample with the base radius of 100$a$ and the height of 10$c$.
Such a shape of the model sample takes into account the demagnetizing field.
We have obtained the values of the hyperfine constants of the transfered hyperfine fields from the fits of the NMR spectra obtained within the PM and UUD phases as shown in Fig.~\ref{fig9_spectra_fits}.
The magnetic moment of the Fe$^{3+}$ ions of the PM phase is 5$\mu_{\rm B}$~\cite{Smirnov_2007}, whereas the ordered moment within the UUD phase is 3.3$\mu_{\rm B}$ at $T=2$~K~\cite{Kenzelmann_2007}.
The details of a numerical modeling of the NMR spectra are described in Ref.~\cite{Sakhratov_2020}.

In our previous work~\cite{Svistov_2006, Sakhratov_2020}, to compute an effective field on the nuclei, we took into account a transfered hyperfine field from the two nearest Fe$^{3+}$ ions of the two neighboring triangular planes.
For rubidium nuclei at (0,~0,~0), the transferred hyperfine field considered was from the two magnetic moments of Fe$^{3+}$ ions located at $(0,~0,~\pm \frac{c}{2})$.
Attempts to fit the experimental spectra of the UUD and PM phases with fixed values of the magnetic moment by only varying the transferred hyperfine field constant $A_1$, were unsuccessful.
For this reason we were forced to also include the additional transferred hyperfine fields from the twelve iron ions belonging to the triangular planes above and below the $^{87}$Rb nucleus, located at $(\pm a,~0,~\pm \frac{c}{2})$, $(0,~\pm a,~\pm \frac{c}{2})$, $(+a, +a, \pm \frac{c}{2})$, and $(-a, -a, \pm \frac{c}{2})$ with a hyperfine constant $A_2$.

The effective transferred hyperfine field for the polarized PM phase is then $(2 A_1 + 12 A_2 ) \times 5\mu_{\rm B}$.
For the UUD phase, there are two inequivalent sites with different fields acting on a rubidium nucleus depending on the nearest Fe$^{3+}$ moments being either parallel or antiparallel to each other.
There are twice as many sites with the antiparallel arrangement of the iron magnetic moments, therefore the NMR spectra within the UUD phase are split into two lines with the intensity ratio 2 to 1.
For the rubidium ions located between the two antiparallel spins, the transferred hyperfine field is $6 A_2 \times 3.3\mu_{\rm B}$, while for another site corresponding to an NMR line with half the intensity, the transfered hyperfine field is $2  A_1 \times 3.3 \mu_{\rm B}$.
By fitting a single NMR line in the  polarized PM phase and two lines in the UUD phase, the following values for the transferred hyperfine fields were obtained, $A_1=-1.22$~mT/$\mu_{\rm B}$ and $A_2=1.31$~mT/$\mu_{\rm B}$.
The blue dashed lines in Fig.~\ref{fig9_spectra_fits} show the NMR spectra computed with these parameters of transfered hyperfine field for the UUD, V and PM phases for a central line.
A small offset  in the $x$-axis is likely to be due to a difference of the studied sample shape from the form assumed in the model or due to the uncertainty in the value of the applied field of a large superconducting solenoid because of frozen fields.
The red lines show the best match to the experimental data when the computed lines are offset by $\approx$~0.008~T.
For all the fits shown in the paper, the field corrections required were less than 0.01~T.

Our neutron diffraction results suggest that in the entire range of temperatures and fields studied, the magnetically ordered phases of \rfm\ are characterized by the propagation vector (1/3, 1/3, $l$).
In this case, the dipolar fields from the Fe$^{3+}$ ions acting on the $^{87}$Rb nuclei are caused only by the components of the magnetic moments parallel to the applied field, as the dipolar fields from the components perpendicular to the field cancel each other.
The NMR spectra in the magnetically ordered states are either broader or more structured than those in a unordered state, suggesting that for all structures in the ordered state there are significant oscillations of the magnetic moments parallel to the applied field.

\begin{figure}[tb]
\includegraphics[width=1.1\columnwidth]{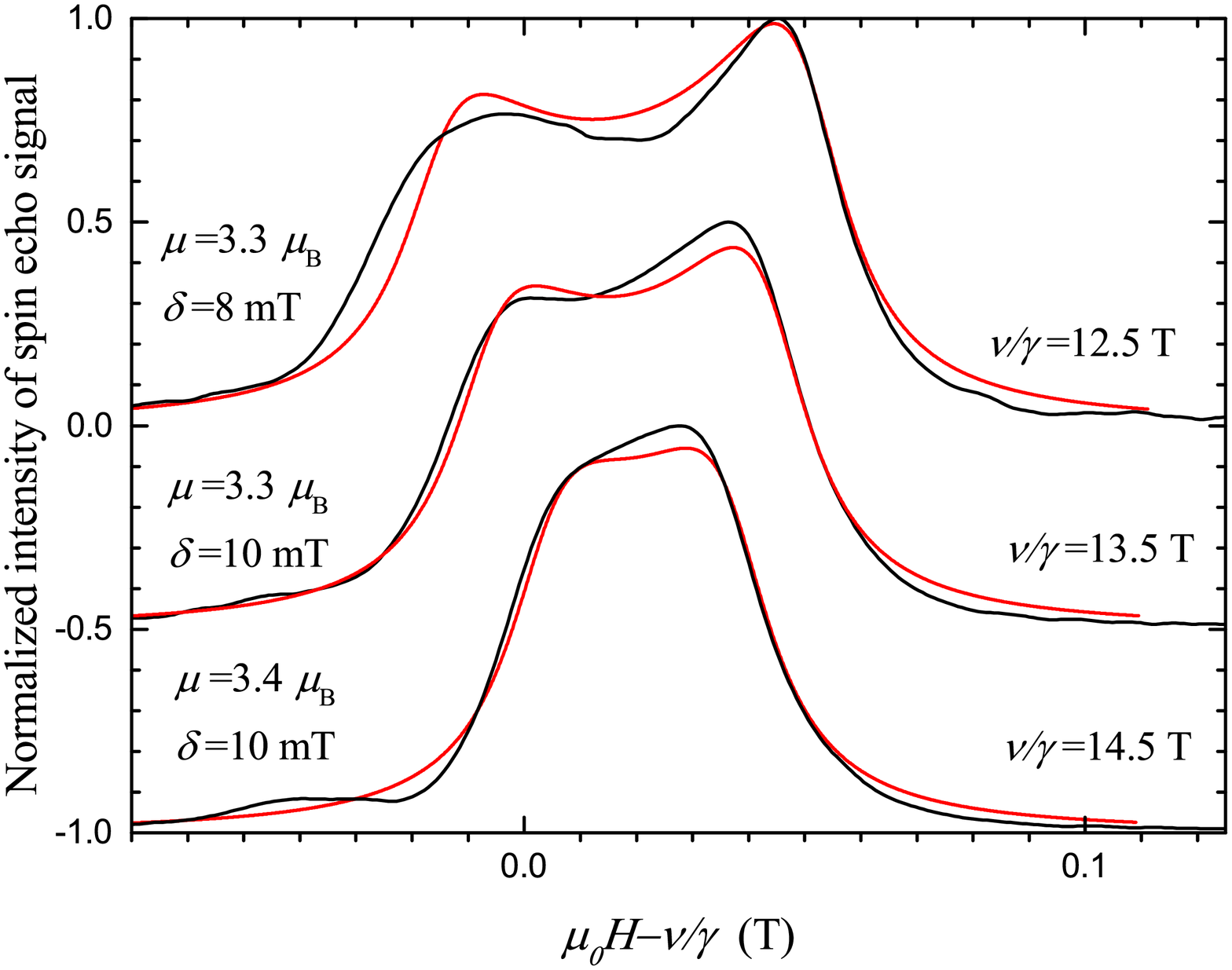}
\caption{$^{87}$Rb NMR spectra measured in the fan phase for a sample from batch II in fields of $\mu_0 H=12.5$, 13.5, and 14.5~T.
The experimental data are shown with black lines while the red lines represent the fits.
The ordered moments of the Fe$^{3+}$ ions, $\mu$, and the width of the individual NMR lines, $\delta$, used for the computation of the spectra are given near the NMR lines.
The spectra are consecutively offset for clarity.}
\label{fig10_spectra_fan}
\end{figure}

For the commensurate Y and V phases, one can expect two-line NMR spectra with an intensity ratio 2 to 1, because the projections of the magnetic moments on the applied field for these structures can be considered as being composed of the UUD and field-polarized phases.
As expected, the NMR spectra measured in the Y, UUD, and V phases have the identical structure (see Fig.~\ref{fig5_NMR_all} and also Ref.~\cite{Svistov_2005}).
We adopted the following procedure for modeling the NMR spectra.
First, using the data for the field dependence of the magnetic moment $M(H)$ from Ref.~\cite{Smirnov_2007}, we determined the magnetic moment directions for the Y and V structures.
For the V phase, as can be seen in Fig.~\ref{fig1_structures}, the angles between the magnetic moments and the field direction are $\Theta_1 = \arccos{\frac{3m^2+1}{4m}}$ and $\Theta_{2,3} = -\arccos{\frac{3m^2-1}{2m}}$
for a single sublattice and for the two aligned sublattices respectively (here $m=M/M_{\rm sat}$).
For the Y phase, the angles are $\Theta_1=0$ and $\Theta_{2,3} = \pm \arccos \frac{3m+1}{2}$.

With the directions of the sublattices magnetic moments determined, the NMR spectra are modeled by the numerical summations of dipolar and transfered hyperfine fields on $^{87}$Rb nuclei using only one free parameter, the value of the ordered magnetic moment on the Fe$^{3+}$ ions.
The width of the NMR line from the nuclei with the equivalent magnetic environment was defined from the line-width of the single-line spectra observed in the PM phase.
The results of the modeling of the NMR spectra in the V phase, in an applied field of $\mu_0 H=10.55$~T, are shown by the blue dashed and red solid lines in Fig.~\ref{fig9_spectra_fits}, where the red line includes a small field correction described above.
The obtained value of the ordered magnetic moment on the iron ion used as a fitting parameter is then $0.8 \times 5\mu_{\rm B}$.

\begin{figure}[tb]
\includegraphics[width=0.95\columnwidth]{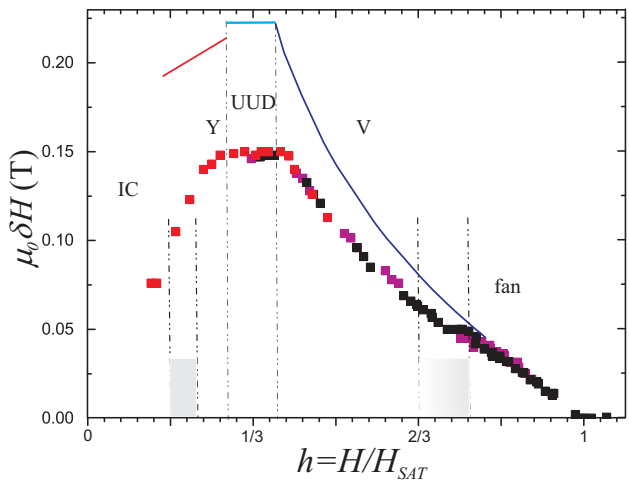}
\caption{Field dependence of the distances between the maxima of the NMR spectra, $\mu_0 \delta H (h)$, measured at $T=1.5$~K on samples of batch~I (red and magenta symbols) and batch~II (black symbols).
$h=H/H_{\rm sat}$.
$\mu_0 H_{\rm sat}=17$ and 16.25~T for samples from batch I and II, respectively.
Red symbols show the results of previous work~\cite{Svistov_2005}.
Solid lines show the $\mu_0 \delta H$ obtained from the NMR spectra computed for the commensurate magnetic structures Y, UUD and V with the magnetic moment of the Fe$^{3+}$ ions $\mu=5\mu_{\rm B}$.
Dash-dotted lines indicate the phases boundaries.
The field regions of mixed magnetic phases are shown in gray. }
\label{fig11_delta_H}
\end{figure}

When modeling the fan phase, as given by Eq.~\ref{eqn_fan}, we first find the value of $L(H)$ from the $M(H)$.
We are then able to find the effective fields acting on the $^{87}$Rb nuclei in different magnetic environments using the value of the ordered magnetic moment of the Fe$^{3+}$ ions as a fitting parameter.
The fitting results for the central NMR line in fields of 12.5, 13.5, and 14.5~T on a sample from batch II are shown in Fig.~\ref{fig10_spectra_fan}.
The values of the ordered component of the magnetic moment on the Fe$^{3+}$ ions corresponding to the best fit are given next to the NMR lines.

\begin{figure}[tb]
\includegraphics[width=0.95\columnwidth]{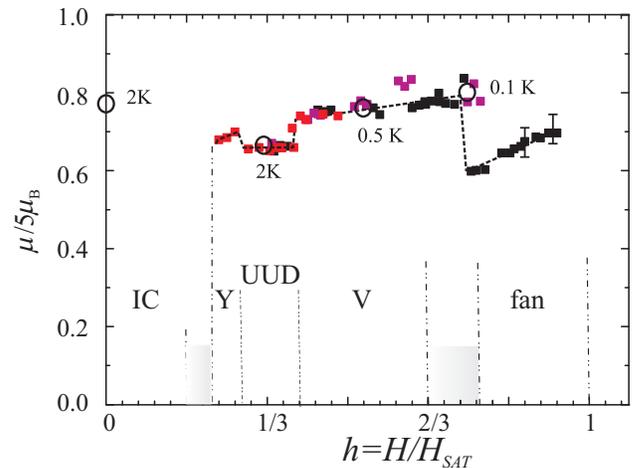}
\caption{Field dependence of the ordered component of the magnetic moment $\mu (h)$ at $T=1.5$~K, $h=H/H_{\rm sat}$.
The values of $\mu$ within the fan phase were obtained from the fits to the NMR spectra as described in the main text.
Different colors show the results obtained on the samples from different batches with the color scheme identical to Fig.~\ref{fig11_delta_H}.
Open circles show the values of the ordered components of magnetic moments from Refs.~\cite{Kenzelmann_2007, White_2013}, with the exact temperature of the neutron diffraction experiments given near the symbols.}
\label{fig12_OrderedM}
\end{figure}

Figure~\ref{fig11_delta_H} shows the field dependence of the distances between the maxima of the NMR spectra, $\mu_0 \delta H$, measured on samples from the two batches.
The data obtained for different samples show a good agreement if plotted as a function of the scaled magnetic field, $h=H/H_{\rm sat}$.
$\mu_0 H_{\rm sat}=17$ and 16.25~T for samples from batch~I and II, respectively.
Red symbols show the results of our previous work~\cite{Svistov_2005}.
Here the data shown are collected from the NMR spectra of all three satellites.
The solid lines show the $\mu_0 \delta H$ obtained from modeling of the NMR spectra for the commensurate magnetic structures Y, UUD and V.
For modeling of the spectra the value of magnetic moments of the three sublattices was equal to $g S \mu_{\rm B}=5\mu_{\rm B}$.
The appreciable difference between the modeling curves and the experimental points could be explained by a further reduction of the ordered component of magnetic moments on the Fe$^{3+}$ ions by thermal or quantum fluctuations.
The field dependence of the value of the ordered components at $T=1.5$~K is shown in Fig.~\ref{fig12_OrderedM}.

The values of ordered moment on the iron ions, $\mu$, within the fan phase were obtained from fitting of the NMR spectra as discussed above.
The different colors show the results obtained on the samples from different batches, with the color scheme identical to the one used in Fig.~\ref{fig11_delta_H}.
Open circles show the values of the ordered components of the magnetic moments from Refs.~\cite{Kenzelmann_2007, White_2013} with the temperatures of neutron diffraction experiments indicated.
As one can see, there is a satisfactory agreement between the results obtained by different techniques.

To summarize this section, the presence of the incommensurate fan structure is detected in the high-field range approaching the saturation field.
The exact nature of the interactions leading to the stabilisation of such a structure is not entirely clear at present.
One possible reason could be the chirality of the crystal structure, which can lead to an inequality of the diagonal inter-plane exchange integrals labeled in Fig.~\ref{fig2_crystal} as $J_a$ and $J_b$.
Another possibility is the presence of a frustrated interaction between the nearest and next nearest triangular planes of an exchange or dipolar nature.
This interaction must provide a very small energy gain in favor of the incommensurate fan phase.
According to a rough estimate discussed in Appendix~\ref{AppB}, the value of the energy gain can be evaluated as $10^{-5}$ to $10^{-4}$ of the strongest in-plane exchange interaction $JS^2$ per Fe$^{3+}$ ion.

\section{Conclusions}

An $^{87}$Rb NMR study of the magnetic phase diagram of \rfm\ within the field range from $H_{\rm sat}/3$ to $H_{\rm sat}$ shows four magnetic phases with distinct NMR spectra.
The modeling of the NMR spectra allows for a verification of the magnetic structures.
It also provides the values of the ordered component of the magnetic Fe$^{3+}$ ions for the whole set of magnetic structures studied.
High field neutron diffraction experiments show a single transition from a magnetically polarized PM structure to the ordered structure with wave vector (1/3, 1/3, 0.43).
A single transition seen in neutron diffraction is in agreement with a single singularity of the temperature or field dependence of the spin-lattice and spin-spin relaxation times of $^{87}$Rb nuclei.
The neutron diffraction experiment at the point of the $H-T$ phase diagram within the high-field phase ($\mu_0 H=14.9$~T and $T=100$~mK) described in Ref.~\cite{White_2013} shows that the components perpendicular to the applied magnetic field are well described by Eq.~\ref{eqn_fan}.
The $^{87}$Rb NMR experiments described here show that the components along the field also are in agreement with the fan structure given by Eq.~\ref{eqn_fan}.
The observation of neutron diffraction reflections corresponding to the second-oder harmonic also indicate the high-field fan phase.


A fan phase is observed in a number of magnetic systems with a two-component order parameter in the field range close to $H_{\rm sat}$~\cite{Chernyshov_2005, Wilson_2013, Lautenschlager_1993, Cemal_2017}.
In the majority of cases the helicity in these systems results from the frustrated interactions between nearest and next nearest ferromagnetically coupled planes, or magnetic ions in the case of quasi one-dimensional magnets.
Fan phases in these magnets results from the competition of helicity and anisotropies~\cite{Nagamia_1961, Gvozdikova_2016,Utesov_2020}.

In the case of \rfm\ the situation seems to be different due to the simultaneous frustration of strong antiferromagnetic interactions within the triangular planes and frustration of inter-plane interactions of an exchange or dipolar nature.
The inter-plane interaction, despite being small, is expected to be responsible for a fan phase in a wide field range below saturation.

\acknowledgments
We thank A.I.~Smirnov, S.S.~Sosin, M.E.~Zhitomirskiy, and J.S.~White for stimulating discussions as well as M.~Bartkowiak for his assistance during the neutron diffraction experiment.
We also gratefully acknowledge the support of the HFM team, S.~Gerischer, R.~Wahle, S.~Kempfer, P.~Heller, and P.~Smeibidl during the HFM/EXED experiment.
Yu.A.S. thanks the Government of the Republic of Tatarstan for the support through the Algarysh Grant.
The modeling of NMR spectra was supported by the Russian Science Foundation Grant No. 17-12-01505.
Work at the National High Magnetic Field Laboratory is supported by the User Collaborative Grants Program (UCGP) under NSF Cooperative Agreement No.~DMR-1157490 and DMR-1644779, and the State of Florida. The work at the University of Tennessee (H.D.Z.) was supported by the NSF with Grant No. NSF-DMR-2003117.

\appendix
\section{Transition to the V$'$ magnetic phase} \label{AppA}

The V~structure for a single triangular plane is defined by the three sublattices $S_1=S_2$, $S_3$ (see Fig.~\ref{fig1_structures}).
For the model of stacked triangular planes with a small antiferromagnetic exchange interaction between the nearest magnetic moments of the neighboring planes ($J'$ in Fig.~\ref{fig2_crystal})  the two expected magnetic structures, V and V$'$, are shown in Fig. ~\ref{fig3_MagStructures}.
The in-plane energies of both structures are identical.
The energy difference between the V and V$'$ structures in this model is solely due to the inter-plane interaction, $\sum\limits_{n=1,3} J' (S_{n,i}S_{n,i+1})/3$.
For the V~phase, the spins of the neighboring planes ($i$ and $i+1$) are obtained by a cyclic permutation of the index $n$ as $S_{n,i+1}=S_{n+1,i}$.
For the V$'$ phase, the components of the spins perpendicular to the applied field alternate from one plane to another as:
\begin{equation}
S'_{n,i+1}=
\begin{pmatrix}
-1 & 0 \\
0 & 1
\end{pmatrix}
\cdot S_{n,i},
\end{equation}
here the $x$-projection is perpendicular to the field.
A simple calculation shows that the V~phase is favorable in fields lower than $H_c=H_{\rm sat}/\sqrt{3}$ while in higher fields, the V$'$~phase becomes favorable.
For the structure at $H_c$, one sublattice ($S_1$ in Fig.~\ref{fig3_MagStructures}) is perpendicular to the applied field.

\section{ High field fan magnetic phase} \label{AppB}

\begin{figure}[tb]
\vspace{5mm}
\includegraphics[width=0.78\columnwidth,angle=0,clip]{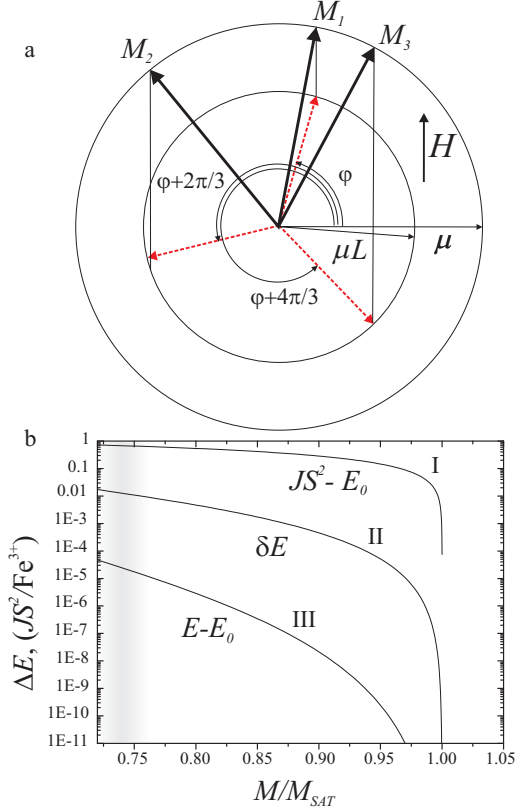}
\caption{(a) Schematic of the magnetic fan structure given by Eq.~(\ref{eqn_fan}).
(b) Energy dependence versus normalized magnetization, $M/M_{\rm sat}$, for
	``I" - the difference between the exchange energy of the PM~phase, $JS^2$, and the exchange energy of the ground state, $E_0$;
	``II" - the variation of the exchange energy from one plane to another for the fan structure, $\delta E$;
	``III" - the difference between the energy of the fan structure, $E$, and the $E_0$.
	The energies were computed in the limit of large spins with the exchange energy given by the first term of Eq.~\ref{fig1_structures}.
	A gray area marks the region of magnetic fields where the transition from the V to fan structure is experimentally observed.}
\label{fig13_deltaE}
\end{figure}

A schematic of the magnetic fan structure given by Eq.~\ref{eqn_fan} is shown in the top panel of Fig.~\ref{fig13_deltaE}.
The arrows $M_{1,2,3}$ show the directions of the magnetic moments of the three sublattices for one triangular plane.
The moment directions change from one plane to another and are defined by the phase $\phi=k_{ic} \, c \, i$, where $k_{ic}$ is the incommensurate wave vector of the structure, $i$ is the index of the triangular plane.
The magnetic moment of every plane is directed along the applied field and its value oscillates from one plane to another around the mean value of $M_0=\sum\limits_{i=1,N}(M_{1,i}+M_{2,i}+M_{3,i})/3N$.

Here, let us consider only the strongest interactions, the in-plane exchange and Zeeman term given by Eq.~\ref{eqn_energy}.
In this case, the energy of the fan structure is definitely higher than the ground state energy of the structure composed from planes with equal magnetic moment $M_0$ corresponding to the energy minimum for every individual plane in a field $H$.
In a quasi-classical limit, $M_0=M_{\rm sat}H/H_{\rm sat}$.
Here $M_{\rm sat}=gS\mu_{\rm B}$ is the magnetic moment of an ion in a PM phase.
Since the mean magnetic moments of the structures compared here are the same, it is sufficient to consider only the exchange energies.
The exchange energy of the ground state magnetic structure computed in the limit of large spin is $E_0=JS^2[\frac{3}{2}(\frac{M_0}{M_{\rm sat}})^2-\frac{1}{2}]$.
This energy grows from $-1/2JS^2$ to $JS^2$ as the field increases from $H=0$ to $H=H_{\rm sat}$.
The value of $E_0$ subtracted from the exchange energy in the saturated phase, $JS^2-E_0$, is given by the curve ``a" in Fig.~\ref{fig13_deltaE}.
Curve ``c" shows the difference between the energies of the fan structure, $E$, and the energy of ground state, $E_0$.
$E$ was found numerically by summing the exchange energies of a large number of triangular planes.
Curve ``b" in the same figure shows the variation of the exchange energy from plane to plane for the fan structure, $\delta E$.
The calculations performed showed that despite the fact that the exchange energy of the fan structure varies significantly from plane to plane, the energy of a large number of planes $E$ in high fields practically coincides with the energy of the ground state $E_0$.
Therefore at the transition point $M\approx 0.75M_{\rm sat}$, the interplane interaction which selects the fan structure provides an energy gain to $E_0$ of the order of only $10^{-5}$  to $10^{-4}$ of $JS^2$ per magnetic ion.

\end{document}